\documentclass[lettersize,journal]{IEEEtran}
\usepackage{amsmath,amsfonts}
\usepackage{algorithmic}
\usepackage{algorithm}
\usepackage{array}
\usepackage[caption=false,font=normalsize,labelfont=sf,textfont=sf]{subfig}
\usepackage{textcomp}
\usepackage{stfloats}
\usepackage{url}
\usepackage{verbatim}
\usepackage{graphicx}
\usepackage{cite}
\usepackage{amssymb}
\usepackage[T1]{fontenc}
\usepackage{array}
\usepackage{booktabs}
\usepackage{multirow}
\usepackage{multicol}
\usepackage{bbding}
\usepackage{hyperref}
\hypersetup{hypertex=true,
	colorlinks=true,
	linkcolor=blue,
	anchorcolor=blue,
	citecolor=blue} 
\begin{document}
	
	\title{Analog-Only Beamforming for Near-Field Multiuser MIMO Communications}
	
	\author{
		Ying Wang,~\IEEEmembership{Graduate Student Member,~IEEE},~and~Chenhao Qi,~\IEEEmembership{Senior Member,~IEEE }
		\thanks{This work was supported in part by the National Natural Science Foundation of China under Grants 62071116 and U22B2007. (Corresponding
			author: Chenhao Qi.)}
		\thanks{Ying Wang and Chenhao Qi are with the School of Information Science and
			Engineering, Southeast University, Nanjing 210096, China (Email: \{wying,
				qch\}@seu.edu.cn).}

	}
	
	\markboth{}%
	{Shell \MakeLowercase{\textit{et al.}}: A Sample Article Using IEEEtran.cls for IEEE Journals}
	
	\IEEEpubid{}
	
	\maketitle
	
	\begin{abstract}
		
		For the existing near-field multiuser communications based on hybrid beamforming (HBF) architectures, high-quality effective channel estimation is required to obtain the channel state information (CSI) for the design of the digital beamformer. To simplify the system reconfiguration and eliminate the pilot overhead required by the effective channel estimation, we considered an analog-only beamforming (AoBF) architecture in this study. The AoBF is designed with the aim of maximizing the sum rate, which is then transformed into a problem, maximizing the power transmitted to the target user equipment (UE) and meanwhile minimizing the power leaked to the other UEs. To solve this problem, we used beam focusing and beam nulling and proposed two AoBF schemes based on the majorization-minimization (MM) algorithm. First, the AoBF scheme based on perfect CSI is proposed, with the focus on the beamforming performance and regardless of the CSI acquisition. Then, the AoBF scheme based on imperfect CSI is proposed, where the low-dimensional imperfect CSI is obtained by beam sweeping based on a near-field codebook. Simulation results demonstrate that the two AoBF schemes can approach the sum rate of the HBF schemes but outperform HBF schemes in terms of energy efficiency (EE).
	\end{abstract}
	
	\section{Introduction}
\IEEEPARstart{M}{illimeter} wave (mmWave) communications and terahertz (THz) communications are regarded as key technologies to support the very high data rate  for future wireless communications. To deal with the serious path loss of signal propagation in these bands, large-scale antenna arrays are equipped at base stations (BSs) to improve the beam gain. Fortunately, the small wavelength of mmWave and THz wave allows the integration of more antennas on a small surface. However, the utilization of large-scale arrays enlarges the Rayleigh distance to dozens of meters, which makes the near-field effect not negligible. Different from far field, the wireless propagation in the near field is precisely modeled as spherical waves instead of plane waves~\cite{ZHY-review, CKJ2}. Thus, the near-field beam can be focused on a specific location, which is different from the far-field beam aligned to a certain angle regardless of distance. Therefore, the conventional far-field multiuser interference suppression methods may not be suitable for near-field multiuser communications~\cite{Limited}.

\begin{figure*}[t]
	\centering	
	\includegraphics[width=6.7 in]{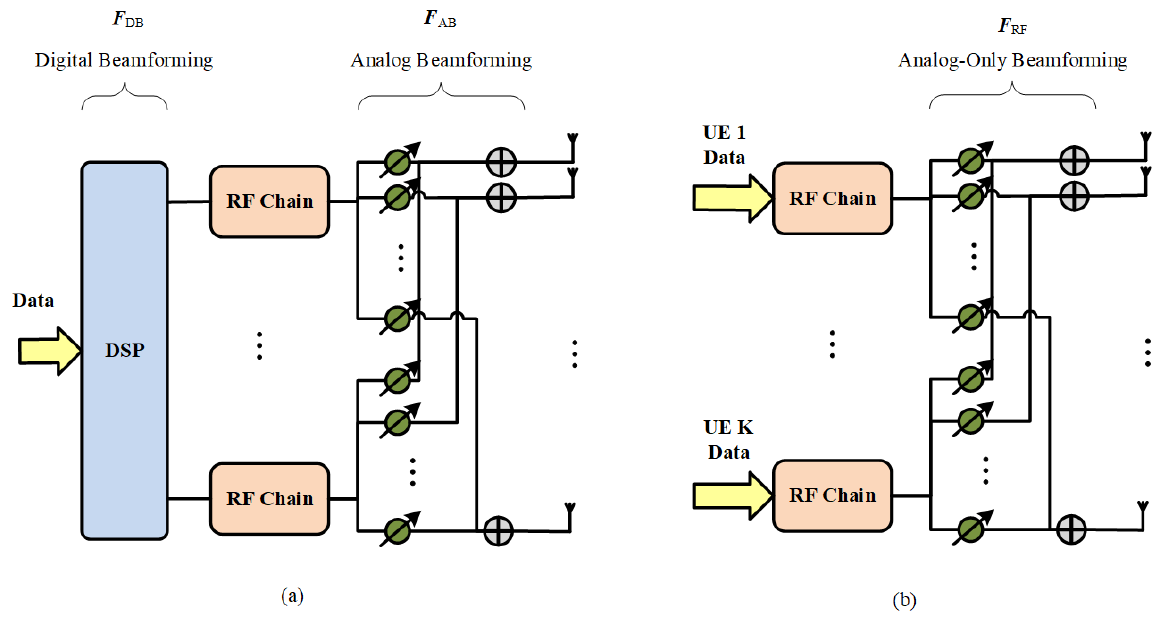}
	\caption{Beamforming architecture comparison: (a) the HBF architecture; (b) the AoBF architecture.}
	\label{fig2}
\end{figure*}

To mitigate the multiuser interference in the near field, various beamforming architectures are designed. The most flexible solution is the fully digital beamforming architecture~\cite{ZHY}. However, this solution causes overwhelming hardware costs for large-scale MIMO systems since each antenna needs to be connected to a radio frequency (RF) chain. To deal with this problem, hybrid beamforming (HBF) architectures are widely used. An HBF scheme for near field is proposed to perform analog beamforming and digital beamforming alternatively until the HBF results approach the solutions of the fully digital beamforming~\cite{ZHY}. 
However, this scheme is designed based on the perfect knowledge of channel state information (CSI), which is impractical.  One method is to estimate the near-field channels~\cite{ununi_codeword}. Due to the equipped large-scale antenna arrays, the dimension of the channel matrix is high , and the channel estimation is complex and challenging. To address this problem, beam sweeping is performed to obtain the low-dimensional imperfect CSI. Based on the imperfect CSI obtained by beam sweeping, a two-stage HBF scheme is proposed for far-field communications~\cite{Limited}. In the first stage of this scheme, the analog beamforming is designed according to the code word selected by beam sweeping, and then, the effective channel~\cite{SXY} is estimated. In the second stage, digital beamforming is designed based on the effective channel to mitigate the multiuser interference. This two-stage HBF method is extended to the near field~\cite{Dai}. To simplify the system reconfiguration and eliminate the pilot overhead required by the effective channel estimation, in this study, we considered using analog-only beamforming (AoBF) architecture to replace the HBF. { Similar to the reconfigurable intelligent surface (RIS) composed of passive reflecting elements~\cite{RIS1, RIS2}, AoBF can flexibly adjust the beam pattern only relying on phase shifters.  As shown in Fig.~\ref{fig2}, the AoBF architecture omits the digital beamforming module. The removal of the digital beamforming module can improve the energy efficiency (EE)~\cite{Constant-Modulus,phase-only-ADMM}. Note that AoBF can be adopted by the BS to simplify the system reconfiguration, while the RIS is typically placed between the BS and the user equipment (UE) to provide additional propagation path for the wireless signal~\cite{RIS4}. Besides, the architectures of AoBF and RIS are also different considering the deployment of RF chains, where the RF chain includes mixers, analog-to-digital converters (ADC) or digital-to-analog converters (DAC), and date converters~\cite{RF}. These components are necessary for the BS but not included in the RIS.}  To reduce the pilot overhead, an AoBF scheme based on the majorization-minimization (MM) method is proposed for partially connected mmWave MIMO in the far field, which is named as A-MM~\cite{HJL}. { However, this far-field scheme cannot be  directly extended to near-field communications due to the inappropriateness of channel modeling and codebook design.

	In this study, we considered the AoBF for near-field multiuser MIMO systems. The AoBF is designed with the aim of maximizing the sum rate, which is then transformed into a problem, maximizing the power transmitted to the target UE and meanwhile minimizing the power leaked to the other UEs. To solve this problem, we used beam focusing and beam nulling and proposed two AoBF schemes based on the MM algorithm. First, the AoBF scheme based on perfect CSI is proposed  to show the feasibility of replacing HBF with AoBF in the near field, with the focus on the beamforming performance and regardless of the CSI acquisition. Then, the AoBF scheme based on imperfect CSI is proposed to meet the practical requirement where the low-dimensional imperfect CSI is obtained by beam sweeping based on a near-field codebook. Since the best code word from beam sweeping corresponds to an area rather than an accurate location, we proposed a heuristic method to approximate the channel vector by generating auxiliary points within the area.}

\textit{Notations}: Symbols for vectors (lower case) and matrices (upper case) are in boldface. The symbols $\mathbb{C}$ and $\mathbb{R}$ denote the set of complex-valued and real-valued numbers, respectively.  For a vector $\boldsymbol{a}$, $\|\boldsymbol{a}\|_2$ and $[\boldsymbol{a}]_n$ denote the $l_2$-norm and the $n$th entry, respectively. For a matrix $\boldsymbol{A}$, $[\boldsymbol{A}]_{m,:}$, $[\boldsymbol{A}]_{:,n}$, $[\boldsymbol{A}]_{m,n}$, and ${\boldsymbol A}^{\rm H}$ denote the $m$th row, the $n$th column, the entry at $m$th row and $n$th column, and the conjugate transpose (Hermitian), respectively. $\Re\{\cdot\}$ denotes the real part of a complex-valued number. $\angle(\cdot)$ denotes the phase of a complex-valued number. $\mathcal{CN}({ 0},\sigma^{2})$ denotes the complex Gaussian distribution with zero mean and variance being $\sigma^{2}$.

\section{System Model}\label{systemmodel}

\begin{figure*}[t]
	
	\centering	
	\includegraphics[width=6.9 in]{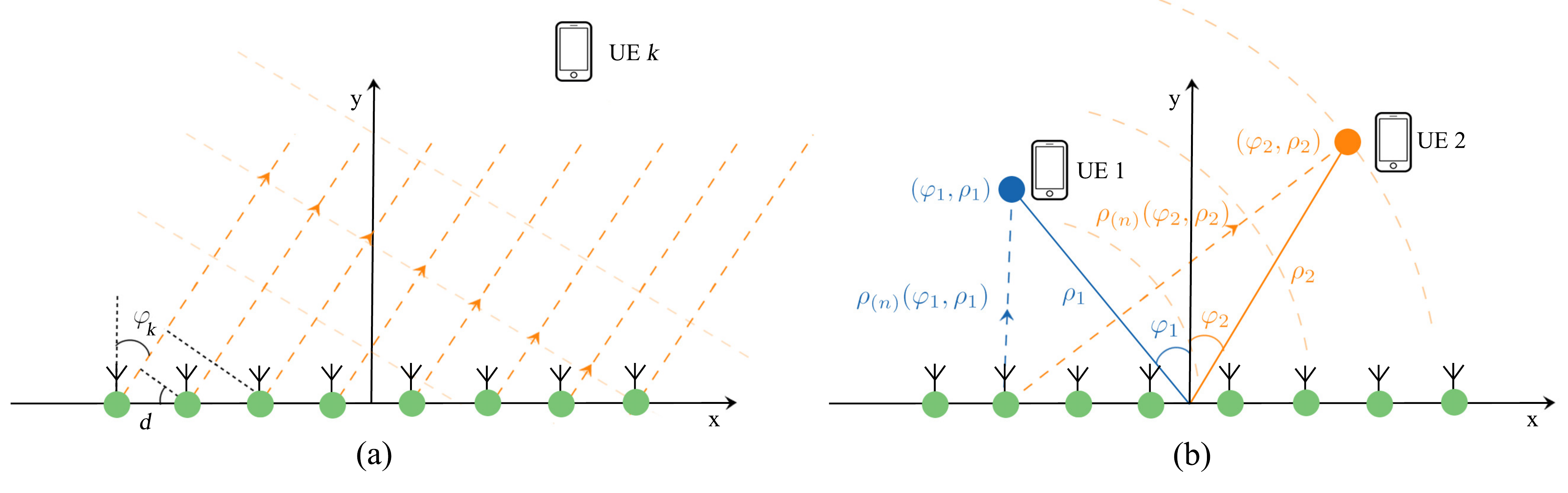}
	\caption{The channel models: (a) in the far field; (b) in the near field.}
	\label{fig1}
\end{figure*}

Considering the downlink transmission for the multiuser MIMO system in the near field, the BS is equipped with $N_{{\rm BS}}$ antennas in terms of uniform linear arrays (ULAs) and serves $K$ UEs simultaneously. The number of UEs served simultaneously is restricted by the number of RF chains at the BS, i.e., $K\leq N_{{\rm RF}}$. For simplicity, we assume the RF chains are fully utilized in this study, i.e., $K= N_{{\rm RF}}$. The origin $(0,0)$ of the XY-plane is assumed to be located at the center of the ULA as shown in Fig.~\ref{fig1}. The location of the $k$th UE is denoted as $(x_k,y_k)$ in Cartesian coordinates. The conversion between Cartesian and polar coordinates can be expressed as follows:
\begin{subequations}
	\label{eq1}
	\begin{align}
		&({\varphi}_{k},\rho_{k})=\left(\arctan\frac{x_k}{y_k},\sqrt{x_k^2+y_k^2}\right) ,\label{eq1a}\\
		&(x_k,y_k)=\left(\rho_{k}\sin\varphi_{k},\rho_{k}\cos\varphi_{k}\right). \label{eq1b}
	\end{align}
\end{subequations}
Given the wavelength $\lambda$ of signals and the adjacent distance between antennas $d=\lambda/2$, the $n$th antenna in ULA is located at $(d\gamma_n,0)$, where $\gamma_n\triangleq n-(N_{{\rm BS}}+1)/2,  n= 1, 2, \dots, N_{{\rm BS}}$. The distance between the $k$th UE and the $n$th antenna can be computed by using the following equation~\cite{far-field}:
\begin{equation}
	\label{eq2}
	\begin{aligned}
		\rho_{(n)}(\varphi_{k},\rho_{k})=&\sqrt{(x_k-d\gamma_n)^2+y_k^2}\\
		=&\sqrt{\rho_{k}^2+d^2\gamma_n^2-2d\gamma_n\rho_{k}\sin\varphi_{k}}.
	\end{aligned}
\end{equation}
Then, the channel steering vector in the near field is defined as follows:
\begin{equation}
	\begin{aligned}
		\label{eq3}
		&\boldsymbol{u}( \varphi_{k},\rho_{k})\triangleq\\
		&\frac{1}{\sqrt{N_{{\rm BS}}}}{\left[e^{\frac{-j2\pi\rho_{(1)}(\varphi_{k},\rho_{k})}{\lambda} },\dots, e^{\frac{-j2\pi\rho_{(N_{{\rm BS}})}(\varphi_{k},\rho_{k})}{\lambda} }\right]^{\rm T}}.
	\end{aligned}
\end{equation}
{As shown in Fig.~\ref{fig1}a, the distance in the far field under planar wave assumption can be simplified as follows~\cite{far-field}:
	\begin{equation}
		\label{eq2b}
		\begin{aligned}
			\rho_{(n)}(\varphi_{k},\rho_{k})
			=&\rho_{k}(1+\frac{d^2\gamma_n^2}{\rho_{k}^2}-\frac{2d\gamma_n\sin\varphi_{k}}{\rho_{k}})^{\frac{1}{2}}\\
			{\approx}&\rho_{k}(1-\frac{d\gamma_n\sin\varphi_{k}}{\rho_{k}})=\rho_{k}-d\gamma_n\sin\varphi_{k},			
		\end{aligned}
	\end{equation}
	where ${(d^2\gamma_n^2)}/{\rho_{k}^2}$ is ignored since $\rho_{k}\gg d$ in the far field, and then, the approximation is performed according to the first-order Taylor expansion $(1 + x)^{1/2}\approx 1 + 1/2x$. By ignoring the constant phase independent of $n$, the channel steering vector in the far field can be denoted as $\boldsymbol{u}'( \varphi_{k})\triangleq \frac{1}{\sqrt{N_{{\rm BS}}}} {\left[e^{{j2\pi d\sin\varphi_{k}}/{\lambda} }, \dots , e^{{j2\pi N_{\rm BS}d\sin\varphi_{k}}/{\lambda} }\right]^{\rm T}}$, whose phases are linear to the antenna index $n$. Note that the far-field steering vector is only related to the angle of the UE, regardless of the distance. In the near field, the assumption $\rho_{k}\gg d$ does not hold anymore, and the approximation based on Taylor expansion is not accurate when $n$ is large. Thus, the planar wave assumption in Eq.~\eqref{eq2b} fails in the near field. On the contrary, the near field steering vector in Eq.~\eqref{eq3} is related to both the angle and the distance. The introduction of the distance dimension provides more degrees of freedom for beamforming.}

Within the Rayleigh distance~\cite{RayleighDistance} $D_{{\rm R}}\triangleq{2N_{{\rm BS}}^2d^2}/{\lambda}$, as shown in Fig.~\ref{fig1}b,  the near-field spherical-wave channel model~\cite{ununi_codeword} for the $k$th UE is denoted as follows:\footnote{In this study, we designed the scheme based on the mathematically abstracted channel modeling for the convenience of signal processing and system analysis~\cite{channel3}. However, the proposed scheme can be extended to the physically consistent modeling (PCM) to consider the physical effects of wave propagation~\cite{channel1,channel2,channel3}. In our future study, we will investigate AoBF based on PCM.}
\begin{equation}
	\label{eq4}
	\boldsymbol{h}_k=\sqrt{\frac{{N}_{{\rm BS}}}{L}}\sum_{l=1}^L {\alpha}_{k,l}{{\boldsymbol{u}}}\left( {\varphi}_{k,l},\rho_{k,l} \right),
\end{equation}
where $L$, ${\alpha}_{k,l}$, and 
$({\varphi}_{k,l},\rho_{k,l})$ are the number of the channel paths, the complex channel gain of the $l$th path, for $l=1,2, \dots, L$, and the polar coordinate of the UE or scatterers, respectively. Given the channel model $\boldsymbol{h}_k$ in  Eq.~\eqref{eq4}, the received signal at the $k$th UE is denoted as follows:
\begin{equation}
	\label{eq5}
	y_k= \sqrt{P} \boldsymbol{h }_k^{\rm H} \boldsymbol{F}\boldsymbol{x}+ {n}_k,
\end{equation}
where $\boldsymbol{F}\in \mathbb{C}^{N_{{\rm BS}}\times K}$,     $\boldsymbol{x}\in \mathbb{C}^{K}$, $y_k$, and ${n}_k$ denote the beamforming matrix at the BS,  the signals transmitted to $K$ UEs, the received signal, and an additive white Gaussian noise (AWGN) with zero mean and $\sigma^2$ variance, i.e., ${n}_k \sim \mathcal{C N}\left({0}, \sigma^2 \right)$, respectively.   In addition, the total power of the BS is $P$, and the transmitted signals subject to the unit power constraint, i.e.,  $\mathbb{E}\{\boldsymbol{x}\boldsymbol{x}^{\rm H}\}=\boldsymbol{I}_K$.

To maximize the sum rate of near-field multiuser MIMO, the beamforming design problem is formulated as follows:
\begin{equation}
	\label{eq6}
	\begin{aligned}
		&\max_{\boldsymbol{F}} \quad \sum\limits_{k=1}^K {R_k},\\
		&\text { s.t. }\quad~ \|\boldsymbol{F}\|_{\rm F}=K,
	\end{aligned}
\end{equation}
where the achievable rate of the $k$th UE can be computed by the following:
\begin{subequations}\label{eq7}
	\begin{align}
		&R_k\triangleq{{\log }_{2}}\left( 1+ {\rm SINR}_k\right),\label{eq7a}\\
		&{\rm SINR}_k\triangleq\frac{\frac{P}{K}{{\left| \boldsymbol{h}_k^{\rm H} \left[\boldsymbol{F}\right]_{:, k} \right|}^{2}}}{\frac{P}{K}\sum\limits_{i\neq k}^{K}{\left|\boldsymbol{h}_k^{\rm H} \left[\boldsymbol{F}\right]_{:, i} \right|}^{2}+{\sigma }^{2}}. \label{eq7b}
	\end{align}
\end{subequations}
Note that ${{\rm SINR}}_k$ represents the signal-to-interference-plus-noise ratio (SINR) for the $k$th UE. Particularly, maximizing the sum rate in Eq.~\eqref{eq6} is equivalent to maximizing the ${\rm SINR}$ of each UE. Thus, we can transform Eq.~\eqref{eq6} into $K$ subproblems as follows:
\begin{equation}
	\label{eq8}
	\begin{aligned}
		&\max_{\boldsymbol{F}} \quad {{\rm SINR}}_k,\\
		&\text { s.t. }\quad~ \big\|\boldsymbol{F}\big\|_{\rm F}=K.
	\end{aligned}
\end{equation}
During the downlink transmission, the BS designs $ \boldsymbol{F}$ to maximize the sum rate. 

For HBF schemes, as shown in Fig.~\ref{fig2}a, the HBF is implemented by the combination of analog beamforming and digital beamforming~\cite{Dai}, i.e.,  $\boldsymbol{F}\triangleq\boldsymbol{F}_{{\rm AB}}\boldsymbol{F}_{{\rm DB}}$. The analog beamforming is denoted as $\boldsymbol{F}_{{\rm AB}}\triangleq [\boldsymbol{v}_{(1)}$, $\boldsymbol{v}_{(2)},\dots,\boldsymbol{v}_{(K)}]$,  where each code word $\boldsymbol{v}_{(k)} \in \mathbb{C}^{N_{{\rm BS}}}$ with the constant modulus constraint $\big|[\boldsymbol{v}_{(k)}]_n\big|=1/\sqrt{N_{{\rm BS}}}$ is determined by beam sweeping and implemented by phase shifters~\cite{Limited}. Then, the effective channel $\widetilde{\boldsymbol{h}}_k=\boldsymbol{F}_{{\rm AB}}^{\rm H}\boldsymbol{h}_k$ can be obtained by pilot-assisted channel estimation. Based on $\widetilde{\boldsymbol{h}}_k$, the digital beamforming $\boldsymbol{F}_{{\rm DB}}$ can be designed via zero forcing (ZF) or weighted minimum mean-squared error (WMMSE) with the constraint $ \big\|\boldsymbol{F}_{{\rm AB}}[\boldsymbol{F}_{{\rm DB}}]_{:,k}\big\|_{\rm F}=1$. This constraint means that the HBF provides no power gain. 

For our AoBF scheme, $\boldsymbol{F}$ is merely designed by analog beamforming, i.e.,  $\boldsymbol{F}\triangleq\boldsymbol{F}_{{\rm RF}}$ as shown in Fig.~\ref{fig2}b. Thus, the pilot overhead required by the effective channel estimation in the HBF schemes can be omitted. The AoBF is denoted as $\boldsymbol{F}_{{\rm RF}} \in \mathbb{C}^{N_{{\rm BS}}\times K}$ with the constant modulus constraint $\left|[\boldsymbol{F}_{{\rm RF}}]_{n,k}\right|=1/\sqrt{N_{{\rm BS}}}$. We design the phase of each entry in $\boldsymbol{F}_{{\rm RF}}$ to maximize the sum rate.

{	\section{AoBF for Near-Field Multiuser MIMO}
	To simplify the system reconfiguration and eliminate the pilot overhead required by the effective channel estimation, we considered the AoBF architecture in this study. In this section, we proposed two AoBF schemes for near-field multiuser MIMO by exploiting the extra degree of freedom in the distance domain provided from spherical waves. The AoBF scheme based on perfect CSI is first proposed to show the feasibility of replacing HBF with AoBF in the near field, with the focus on the beamforming performance and regardless of the CSI acquisition.  Then, the AoBF scheme based on imperfect CSI is designed to meet the practical requirement where the low-dimensional imperfect CSI is obtained by beam sweeping based on a near-field codebook.}

\subsection{AoBF Based on Perfect CSI }
With the focus on the beamforming performance and regardless of the CSI acquisition, we assume that the CSI is perfectly known. As the analog beamforming for all UEs is coupled in Eq.~\eqref{eq7b}, signal-to-leakage-and-noise ratio (SLNR) is applied to replace SINR for decoupling in this study~\cite{MM-SLNR}. The expression of the SLNR in AoBF is denoted as follows:
\begin{equation}\label{eq9}
	{\rm SLNR}_k\triangleq\frac{\frac{P}{K}{{\left| \boldsymbol{h}_k^{\rm H} \left[\boldsymbol{F}_{{\rm RF}}\right]_{:, k} \right|}^{2}}}{\frac{P}{K}\sum\limits_{i\neq k}^{K}{\left|\boldsymbol{h}_i^{\rm H} \left[\boldsymbol{F}_{{\rm RF}}\right]_{:, k} \right|}^{2}+{\sigma }^{2}}.
\end{equation}
Note that the difference between the SINR in Eq.~\eqref{eq7b} and the SLNR in Eq.~\eqref{eq9} lies in the interference term in the denominator. For ${\rm SINR}_k$, ${\sum\limits_{i\neq k}^{K}\left|\boldsymbol{h}_k^{\rm H} \left[\boldsymbol{F}_{{\rm RF}}\right]_{:, i} \right|}^{2}$ denotes the power of the interference from all other UEs to the target UE, i.e., the $k$th UE. For ${\rm SLNR}_k$, ${\sum\limits_{i\neq k}^{K}\left|\boldsymbol{h}_i^{\rm H} \left[\boldsymbol{F}_{{\rm RF}}\right]_{:, k} \right|}^{2}$ represents the power leaked from the target UE to all the other UEs. Reducing the interference power from the other UEs to the target UE and mitigating the leaked power from the target UE to the other UEs are intrinsically equivalent for multiuser interference suppression. Thus, the optimization problem in Eq.~\eqref{eq8} can be converted into the following:
\begin{equation}
	\label{eq10}
	\begin{aligned}
		&\max_{[\boldsymbol{F}_{{\rm RF}}]_{:,k}} \quad {{\rm SLNR}}_k,\\
		&~\text { s.t. }\quad ~~\big|[\boldsymbol{F}_{{\rm RF}}]_{n,k}\big|=\frac{1}{\sqrt{N_{{\rm BS}}}}.
	\end{aligned}
\end{equation} 
To simplify Eq.~\eqref{eq10}, a weight parameter $\omega$ is introduced to the optimization, and the constant term $\omega\sigma^2 K/P $ is removed for its independence of $[\boldsymbol{F}_{{\rm RF}}]_{:,k}$. The objective function of the simplified problem can be expressed as follows:
\begin{equation}
	\label{eq11}
	\begin{aligned}
		&\min_{[\boldsymbol{F}_{{\rm RF}}]_{:,k}} \quad -{\left| \boldsymbol{h}_k^{\rm H} \left[\boldsymbol{F}_{{\rm RF}}\right]_{:, k} \right|}^{2}+ \omega \sum\limits_{i\neq k}^{K}{\left|\boldsymbol{h}_i^{\rm H} \left[\boldsymbol{F}_{{\rm RF}}\right]_{:, k} \right|}^{2}.
	\end{aligned}
\end{equation} 
The original problem in Eq.~\eqref{eq6} is transformed into the problem in Eq.~\eqref{eq11}, maximizing the power focused on the target UE and meanwhile minimizing the power leaked to the other UEs. To solve this nonconvex problem, we use beam focusing and beam nulling based on the MM method~\cite{MM-SLNR}. The MM method is an iterative algorithm, with each iteration consisting of a majorization stage and a minimization stage. In the majorization stage, the original non-convex objective function is transformed into a surrogate function that is easy to optimize at the feasible point. The feasible point is obtained from the initialization or the last iteration. This surrogate function serves as an upper-bound function with the same value and derivative as the original function at the feasible point. Then, in the minimization stage, the surrogate function is minimized to update the feasible point. Owing to the closed-form solutions of the surrogate functions, the iterations achieve fast convergence~\cite{MM2, MM3,MM-SLNR}.

Since the first term in Eq.~\eqref{eq11} is concave, we can get	the following equation:	
\begin{equation}
	\label{eq12}
	\begin{aligned}
		&~~~-{\left| \boldsymbol{h}_k^{\rm H} \left[\boldsymbol{F}_{{\rm RF}}\right]_{:, k} \right|}^{2}\\
		&=-\left[\boldsymbol{F}_{{\rm RF}}\right]_{:, k}^{\rm H } \boldsymbol{h}_k \boldsymbol{h}_k^{\rm H} \left[\boldsymbol{F}_{{\rm RF}}\right]_{:, k}\\
		&\leq -(\left[\boldsymbol{F}_{{\rm RF}}\right]_{:, k}^{(t)})^{\rm H } \boldsymbol{h}_k  \boldsymbol{h}_k^{\rm H} \left[\boldsymbol{F}_{{\rm RF}}\right]_{:, k}^{(t)}\\
		&~~~-2\left(\boldsymbol{h}_k \boldsymbol{h}_k^{\rm H}\left[\boldsymbol{F}_{{\rm RF}}\right]_{:, k}^{(t)}\right)^{\rm H}\left(\left[\boldsymbol{F}_{{\rm RF}}\right]_{:, k}-\left[\boldsymbol{F}_{{\rm RF}}\right]_{:, k}^{(t)}\right)\\
		&=-2\left(\boldsymbol{h}_k  \boldsymbol{h}_k^{\rm H} \left[\boldsymbol{F}_{{\rm RF}}\right]_{:, k}^{(t)}\right)^{\rm H}\left[\boldsymbol{F}_{{\rm RF}}\right]_{:, k}+c_1,
	\end{aligned}
\end{equation}
where $\left[\boldsymbol{F}_{{\rm RF}}\right]_{:, k}^{(t)}$ is the feasible point obtained in the $t$th iteration and $c_1$ is a constant independent of $\left[\boldsymbol{F}_{{\rm RF}}\right]_{:, k}$. 
{Since ${\left| \boldsymbol{h}_k^{\rm H} \left[\boldsymbol{F}_{{\rm RF}}\right]_{:, k} \right|}^{2}$ is real-value, we can get $\Re\{-{\left| \boldsymbol{h}_k^{\rm H} \left[\boldsymbol{F}_{{\rm RF}}\right]_{:, k} \right|}^{2}\}=-{\left| \boldsymbol{h}_k^{\rm H} \left[\boldsymbol{F}_{{\rm RF}}\right]_{:, k} \right|}^{2}$. Thus, if we obtain the real value of both sides in Eq.~\eqref{eq12}, we can get the following equation:
	\begin{equation}	
		\label{12a}			
		-{\left| \boldsymbol{h}_k^{\rm H} \left[\boldsymbol{F}_{{\rm RF}}\right]_{:, k} \right|}^{2}\leq -2\Re\big\{\left[\boldsymbol{F}_{{\rm RF}}\right]_{:, k}^{\rm H}\boldsymbol{h}_k  \boldsymbol{h}_k^{\rm H} \left[\boldsymbol{F}_{{\rm RF}}\right]_{:, k}^{(t)} \big\}+c_1,
	\end{equation}
	where 
	\begin{equation}
		\begin{aligned}
			\label{12b}
			&\Re\left\{\left(\boldsymbol{h}_k  \boldsymbol{h}_k^{\rm H} \left[\boldsymbol{F}_{{\rm RF}}\right]_{:, k}^{(t)}\right)^{\rm H}\left[\boldsymbol{F}_{{\rm RF}}\right]_{:, k} \right\}\\
			=&\Re\big\{\left[\boldsymbol{F}_{{\rm RF}}\right]_{:, k}^{\rm H}\boldsymbol{h}_k  \boldsymbol{h}_k^{\rm H} \left[\boldsymbol{F}_{{\rm RF}}\right]_{:, k}^{(t)} \big\}.
		\end{aligned}
\end{equation}}Based on the \textit{Lemma 1} proposed by \cite{MMLemma1}, each entry of the second term in Eq.~\eqref{eq11} satisfies as follows:
\begin{equation}
	\label{eq13}
	\begin{aligned}
		&~~~{\left| \boldsymbol{h}_i^{\rm H} \left[\boldsymbol{F}_{{\rm RF}}\right]_{:, k} \right|}^{2}\\
		&=\left[\boldsymbol{F}_{{\rm RF}}\right]_{:, k}^{\rm H } \boldsymbol{h}_i  \boldsymbol{h}_i^{\rm H} \left[\boldsymbol{F}_{{\rm RF}}\right]_{:, k}\\
		&\leq \left(\left[\boldsymbol{F}_{{\rm RF}}\right]_{:, k}^{(t)}\right)^{\rm H } \boldsymbol{h}_i \boldsymbol{h}_i^{\rm H} \left[\boldsymbol{F}_{{\rm RF}}\right]_{:, k}^{(t)}\\
		&~~~+\mu\left(\left[\boldsymbol{F}_{{\rm RF}}\right]_{:, k}-\left[\boldsymbol{F}_{{\rm RF}}\right]_{:, k}^{(t)}\right)^{H}\left(\left[\boldsymbol{F}_{{\rm RF}}\right]_{:, k}-\left[\boldsymbol{F}_{{\rm RF}}\right]_{:, k}^{(t)}\right)\\ &~~~+2\Re\left\{\left(\left[\boldsymbol{F}_{{\rm RF}}\right]_{:, k}-\left[\boldsymbol{F}_{{\rm RF}}\right]_{:, k}^{(t)}\right)^{H}\boldsymbol{h}_i\boldsymbol{h}_i^{\rm H}\left[\boldsymbol{F}_{{\rm RF}}\right]_{:, k}^{(t)}\right\}\\
		&=\mu\left[\boldsymbol{F}_{{\rm RF}}\right]_{:, k}^{\rm H } \left[\boldsymbol{F}_{{\rm RF}}\right]_{:, k}\\
		&~~~+\left(\left[\boldsymbol{F}_{{\rm RF}}\right]_{:, k}^{(t)}\right)^{\rm H }\left(\mu\boldsymbol{I}_{N_{{\rm BS}}}-\boldsymbol{h}_i\boldsymbol{h}_i^{\rm H}  \right)\left[\boldsymbol{F}_{{\rm RF}}\right]_{:, k}^{(t)}\\
		&~~~+2\Re\left\{\left[\boldsymbol{F}_{{\rm RF}}\right]_{:, k}^{\rm H}\left(\boldsymbol{h}_i\boldsymbol{h}_i^{\rm H}-\mu\boldsymbol{I}_{N_{{\rm BS}}}\right) \left[\boldsymbol{F}_{{\rm RF}}\right]_{:, k}^{(t)} \right\}\\
		&=2\Re\left\{\left[\boldsymbol{F}_{{\rm RF}}\right]_{:, k}^{\rm H}\left(\boldsymbol{h}_i\boldsymbol{h}_i^{\rm H}-\mu\boldsymbol{I}_{N_{{\rm BS}}}\right) \left[\boldsymbol{F}_{{\rm RF}}\right]_{:, k}^{(t)} \right\}+c_2,
	\end{aligned}
\end{equation}
where $c_2$, $\boldsymbol{I}_{N_{{\rm BS}}}\in \mathbb{C}^{N_{{\rm BS}}\times N_{{\rm BS}}}$, and $\mu$ denote a constant independent of $\left[\boldsymbol{F}_{{\rm RF}}\right]_{:, k}$, a unit diagonal matrix, and the maximum eigenvalue of the one-rank matrix $\boldsymbol{h}_i\boldsymbol{h}_i^{\rm H}$, respectively. {Thus, we can get the following equation:
	\begin{equation}
		\label{13a}
		\begin{aligned}
			&~~~{\sum\limits_{i\neq k}^{K}\left| \boldsymbol{h}_i^{\rm H} \left[\boldsymbol{F}_{{\rm RF}}\right]_{:, k} \right|}^{2}\\
			&\leq 2\Re\left\{\left[\boldsymbol{F}_{{\rm RF}}\right]_{:, k}^{\rm H}\sum\limits_{i\neq k}^{K}\left(\boldsymbol{h}_i\boldsymbol{h}_i^{\rm H}-\mu\boldsymbol{I}_{N_{{\rm BS}}}\right) \left[\boldsymbol{F}_{{\rm RF}}\right]_{:, k}^{(t)} \right\}+c_3,
		\end{aligned}
	\end{equation}
	where $c_3\triangleq(K-1)c_2$. By adding the corresponding sides of Eqs~\eqref{12a} and \eqref{13a}, we have the following equation:
	\begin{equation}
		\label{14a}
		\begin{aligned}
			&-{\left| \boldsymbol{h}_k^{\rm H} \left[\boldsymbol{F}_{{\rm RF}}\right]_{:, k} \right|}^{2}+ \omega \sum\limits_{i\neq k}^{K}{\left|\boldsymbol{h}_i^{\rm H} \left[\boldsymbol{F}_{{\rm RF}}\right]_{:, k} \right|}^{2}\\
			\leq&2\Re\left\{[\boldsymbol{F}_{{\rm RF}}]_{:,k}^{\rm H} \left(-\boldsymbol{\eta}_1^{(t)}+\omega\boldsymbol{\eta}_2^{(t)}  \right)\left[\boldsymbol{F}_{{\rm RF}}\right]_{:, k}^{(t)}\right\}+c_4,
		\end{aligned}
	\end{equation}
	where 
	\begin{subequations}
		\label{eq15}
		\begin{align}
			&\boldsymbol{\eta}_1^{(t)}\triangleq\boldsymbol{h}_k  \boldsymbol{h}_k^{\rm H} ,\label{eq15a}\\
			&\boldsymbol{\eta}_2^{(t)}\triangleq\sum\limits_{i\neq k}^{K}\left(\boldsymbol{h}_i\boldsymbol{h}_i^{\rm H}-\mu\boldsymbol{I}_{N_{{\rm BS}}}\right) ,\label{eq15b}\\
			&c_4\triangleq c_1+c_3.\label{eq15c}
		\end{align}
	\end{subequations}The right part of the inequality in Eq.~\eqref{14a} is an available surrogate function of the original objective function in Eq.~\eqref{eq11}. By ignoring the constant variable $c_4$ and the multiplier $2$, the problem in Eq.~\eqref{eq11} can be rewritten as follows:
	\begin{equation}
		\label{eq14}
		\begin{aligned}
			&\min_{[\boldsymbol{F}_{{\rm RF}}]_{:,k}} \quad \Re\left\{[\boldsymbol{F}_{{\rm RF}}]_{:,k}^{\rm H} \left(-\boldsymbol{\eta}_1^{(t)}+\omega\boldsymbol{\eta}_2^{(t)}  \right)\left[\boldsymbol{F}_{{\rm RF}}\right]_{:, k}^{(t)}\right\}.			
		\end{aligned}
	\end{equation}
	Based on $\|\boldsymbol{a}-\boldsymbol{b}\|_{\rm F}^2=\|\boldsymbol{a}\|^2_{\rm F}+\|\boldsymbol{b}\|^2_{\rm F}-2\Re\{\boldsymbol{a}^{\rm H}\boldsymbol{b}\}$, the problem in Eq.~\eqref{eq14} can be rewritten as follows:
	\begin{equation}
		\label{14b}
		\begin{aligned}
			&\max_{[\boldsymbol{F}_{{\rm RF}}]_{:,k}} \quad \left\|[\boldsymbol{F}_{{\rm RF}}]_{:,k}^{\rm H}+ \left(\boldsymbol{\eta}_1^{(t)}-\omega\boldsymbol{\eta}_2^{(t)}  \right)\left[\boldsymbol{F}_{{\rm RF}}\right]_{:, k}^{(t)}\right\|^2_{\rm F}.			
		\end{aligned}
	\end{equation}
	Due to the constant modulus constraint and the property of vectors, the closed-form solution of Eq.~\eqref{14b} is expressed as follows:
	\begin{equation}\label{eq16}
		\left[\boldsymbol{F}_{{\rm RF}}\right]_{n, k}^{(t+1)}= e^{j\angle\left(\left[\left(\boldsymbol{\eta}_i^{(t)}-\omega\boldsymbol{\eta}_2^{(t)}\right)\left[\boldsymbol{F}_{{\rm RF}}\right]_{:, k}^{(t)}\right]_n\right)}.
\end{equation}}The details of the AoBF scheme based on perfect CSI are exhibited in {\bf Algorithm \ref{alg1}}. The \textit{stop condition} in Step~\ref{stop} is set as $\big\|[\boldsymbol{F}_{{\rm RF}}]_{:,k}^{(t)}-[\boldsymbol{F}_{{\rm RF}}]_{:,k}^{(t-1)}\big\|_2^2\leq \epsilon$ or $t>T_{{\rm max}}$.

\begin{algorithm}[!t]
	\caption{AoBF Based on Perfect CSI}
	\label{alg1}
	\begin{algorithmic}[1]
		\STATE \textbf{Input:} $N_{{\rm BS}}$, $\boldsymbol{h}_k$, $\omega$, $\epsilon$, $T_{{\rm max}}$.
		\FOR{$k=1:K$}
		\FOR{$n=1:N_{{\rm BS}}$}
		\STATE Initialize $ \left[\boldsymbol{F}_{{\rm RF}}\right]_{n, k}^{(t)}=\frac{1}{\sqrt{N_{{\rm BS}}}}e^{j \angle\left([\boldsymbol{h}_k]_n\right)}$, $t=0$.
		\ENDFOR
		\REPEAT
		\STATE Obtain $ {\boldsymbol{\eta}}^{(t)}_{1} $ via Eq.~\eqref{eq15a}.
		\STATE Obtain ${ \boldsymbol{\eta}}^{(t)}_{2} $ via Eq.~\eqref{eq15b}.
		\STATE Obtain $\left[\boldsymbol{F}_{{\rm RF}}\right]_{:, k}^{(t+1)}$ via Eq.~\eqref{eq16}.
		\STATE $ t = t + 1 $.
		\UNTIL \textit{stop condition}  is satisfied. \label{stop}
		\STATE $\left[\boldsymbol{F}_{{\rm RF}}\right]_{:, k}= \left[\boldsymbol{F}_{{\rm RF}}\right]_{:, k}^{(t)} $.
		\ENDFOR
		\STATE \textbf{Output:} $\boldsymbol{F}_{\rm RF}$.
	\end{algorithmic}
\end{algorithm}	


\subsection{AoBF Based on Imperfect CSI }

\begin{figure}[t]
	\centering	
	\includegraphics[width=3.4 in]{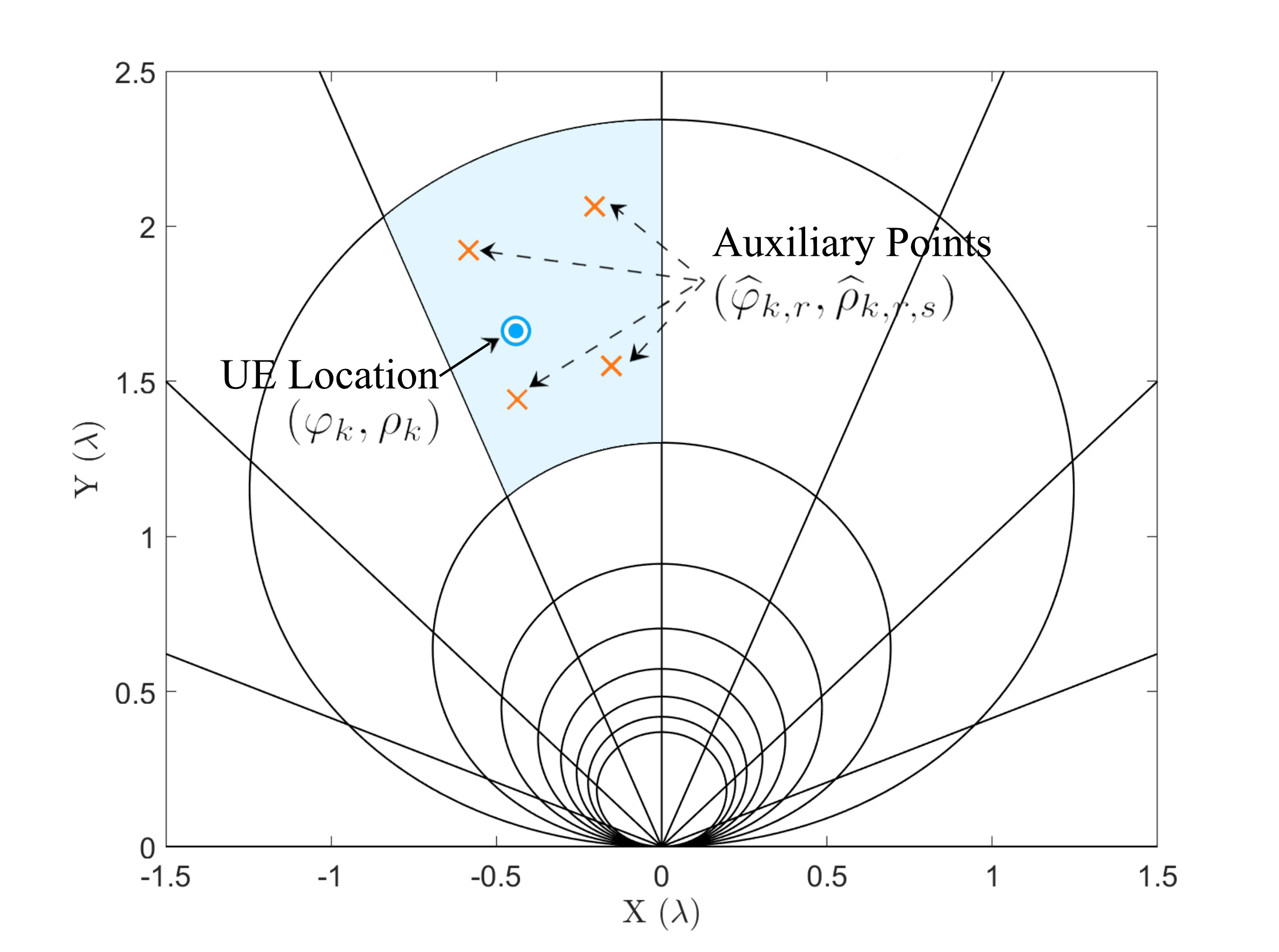}
	\caption{Illustration of beam sweeping code words and auxiliary points.}
	\label{fig3}
\end{figure}

In practice, the perfect CSI is unknown, and the estimation of the high-dimensional CSI is challenging. Normally, to address this problem, beam sweeping is performed to obtain the low-dimensional imperfect CSI~\cite{avoidestimation}. For the near-field beam sweeping, the space is scanned in both the angle and distance dimensions based on a polar-domain near-field codebook $\boldsymbol{\mathcal{V}}$~\cite{ununi_codeword}. The $(p,q)$th code word of $\boldsymbol{\mathcal{V}}$ is given by the following equation:
\begin{equation}\label{eq17}
	\begin{aligned}
		&\boldsymbol{v}_{p,q}\triangleq\boldsymbol{u}(\bar{\varphi}_{p},\bar{\rho}_{p,q}),\\
		&\bar{\varphi}_{p}\triangleq\arcsin\left(\frac{2p-1}{N_{{\rm BS}}}-1\right), ~~p=1, 2,\dots, N_{{\rm BS}},\\
		&\bar{\rho}_{p,q}\triangleq\frac{N_{{\rm BS}}^2d^2}{2q\beta_{\Delta}^2\lambda}\left\{1-\sin^2\varphi_{p}\right\},  ~~q=1, 2,\dots, N_{{\rm DIS}},
	\end{aligned}
\end{equation}
where $p$ and $q$ are the indices in angle and distance dimensions and $\beta_{\Delta}$ is a parameter related to the correlation between adjacent code words at the same angle~\cite{ununi_codeword}.

Since the best code word selected by near-field beam sweeping corresponds to an area rather than an accurate location, we proposed a heuristic method to approximate the unknown $\boldsymbol{h}_k$ by generating auxiliary points within the area as shown in Fig.~\ref{fig3}. The numbers of auxiliary points in the angle and distance dimension are denoted as $R$ and $S$, respectively. The index of the best code word selected by beam sweeping for the $k$th UE is denoted as $(\widetilde{p}_k,\widetilde{q}_k)$. Thus, the angle of the $(r,s)$th auxiliary point in the corresponding area is denoted as follows:
\begin{equation}\label{eq18}
	\begin{aligned}
		&\widehat{\varphi}_{k,r}\triangleq\arcsin\left(\frac{2\widehat{p}_{k,r}-1}{R N_{{\rm BS}}}-1\right),\\ 
		&\widehat{p}_{k,r}\triangleq R(\widetilde{p}_k-1)+r,\\
		&r=1, 2,\dots, R.\\
	\end{aligned}
\end{equation}
\begin{algorithm}[!t]
	\caption{AoBF Based on Imperfect CSI}
	\label{alg2}
	\begin{algorithmic}[1]
		\STATE \textbf{Input:} $N_{{\rm BS}}$, $\widetilde{p}_k$, $\widetilde{q}_k$, $R$, $S$, $\omega$, $\epsilon$, $T_{{\rm max}}$.
		\FOR{$k=1:K$}
		\STATE Initialize $ \left[\boldsymbol{F}_{{\rm RF}}\right]_{:, k}^{(t)}= \boldsymbol{v}_{\widetilde{p}_k,\widetilde{q}_k}$, $t=0$.
		\FOR{$r=1:R$}
		\FOR{$s=1:S$}
		\STATE Obtain $(\widehat{\varphi}_{k,r}, \widehat{\rho}_{k,r,s})$ via Eqs~\eqref{eq18} and \eqref{eq19}.
		\ENDFOR
		\ENDFOR
		\REPEAT
		\STATE Obtain $ {\boldsymbol{\zeta}}^{(t)}_{1} $ via Eq.~\eqref{eq21a}.
		\STATE Obtain ${ \boldsymbol{\zeta}}^{(t)}_{2} $ via Eq.~\eqref{eq21b}.
		\STATE Obtain $\left[\boldsymbol{F}_{{\rm RF}}\right]_{:, k}^{(t+1)}$ via Eq.~\eqref{eq22}.
		\STATE $ t = t + 1 $.
		\UNTIL \textit{stop condition}  is satisfied. \label{stopcondition}
		\STATE $\left[\boldsymbol{F}_{{\rm RF}}\right]_{:, k}= \left[\boldsymbol{F}_{{\rm RF}}\right]_{:, k}^{(t)} $.
		\ENDFOR
		\STATE \textbf{Output:} $\boldsymbol{F}_{\rm RF}$.
	\end{algorithmic}
\end{algorithm}
The distance of the $(r,s)$th auxiliary point is denoted as follows:
\begin{equation}\label{eq19}
	\begin{aligned}
		&\widehat{\rho}_{k,r,s}\triangleq\frac{N_{{\rm BS}}^2d^2}{2{\widehat{q}_{k,s}}\beta_{\Delta}^2\lambda}\left(1-\sin^2\widehat{\varphi}_{k,r}\right),\\
		&\widehat{q}_{k,s}\triangleq\frac{4S}{2S\left(\frac{1}{\widetilde{q}_k}+\frac{1}{\widetilde{q}_k+1}\right)+\left(\frac{1}{\widetilde{q}_k-1}-\frac{1}{\widetilde{q}_k+1}\right)\left(2s-1\right)},\\
		&s=1,2,\dots, S.
	\end{aligned}
\end{equation}
Then, the steering vectors corresponding to the auxiliary points within the area are used to approximate $\boldsymbol{h}_k$, and the objective function in Eq.~\eqref{eq10} can be transformed into the following equation:
\begin{equation}
	\label{eq20}
	\begin{aligned}
		&\max_{[\boldsymbol{F}_{{\rm RF}}]_{:,k}}  \frac{\frac{1}{RS}{{\sum\limits_{r=1}^R\sum\limits_{s=1}^S\left| \boldsymbol{u}(\widehat{\varphi}_{k,r},\widehat{\rho}_{k,r,s})^{\rm H} \left[\boldsymbol{F}_{{\rm RF}}\right]_{:, k} \right|}^{2}}}{\frac{1}{RS}\sum\limits_{i\neq k}^{K}{\sum\limits_{r=1}^R\sum\limits_{s=1}^S\left|\boldsymbol{u}(\widehat{\varphi}_{i,r},\widehat{\rho}_{i,r,s})^{\rm H} \left[\boldsymbol{F}_{{\rm RF}}\right]_{:, k} \right|}^{2}+{\sigma }^{2}}.			
	\end{aligned}
\end{equation} 
Note that the generation of auxiliary points  introduces no extra overhead to the scheme since it is a pure calculation process rather than a sampling operation in the space. Then, the MM algorithm is used to solve Eq.~\eqref{eq20}. The closed-form solution in the $t$th iteration of MM optimization can be denoted as follows:
\begin{equation}\label{eq22}
	\left[\boldsymbol{F}_{{\rm RF}}\right]_{n, k}^{(t+1)}= e^{j\angle\left(\left[\left({\boldsymbol{\zeta}}_i^{(t)}-\omega{\boldsymbol{\zeta}}_2^{(t)}\right)\left[\boldsymbol{F}_{{\rm RF}}\right]_{:, k}^{(t)}\right]_n\right)},
\end{equation}
where
\begin{subequations}
	\label{eq21}
	\begin{align}
		&{\boldsymbol{\zeta}}_1^{(t)}\triangleq\sum\limits_{r=1}^R\sum\limits_{s=1}^S \boldsymbol{u}(\widehat{\varphi}_{k,r},\widehat{\rho}_{k,r,s})  \boldsymbol{u}(\widehat{\varphi}_{k,r},\widehat{\rho}_{k,r,s}) ^{\rm H} ,\label{eq21a}\\
		&{\boldsymbol{\zeta}}_2^{(t)}\triangleq\sum\limits_{i\neq k}^{K}\sum\limits_{r=1}^R\sum\limits_{s=1}^S\left\{\boldsymbol{u}(\widehat{\varphi}_{i,r},\widehat{\rho}_{i,r,s})  \boldsymbol{u}(\widehat{\varphi}_{i,r},\widehat{\rho}_{i,r,s}) ^{\rm H}\nonumber\right.\\
		&\left.~~~~~~~~~~~~~~~~~~~~~~~~~~~-\frac{1}{N_{\rm BS}}\boldsymbol{I}_{N_{{\rm BS}}}\right\} .\label{eq21b}
	\end{align}
\end{subequations}

The AoBF scheme based on imperfect CSI is summarized in {\bf Algorithm \ref{alg2}}. The \textit{stop condition} in Step \ref{stopcondition} is set as $\big\|[\boldsymbol{F}_{{\rm RF}}]_{:,k}^{(t)}-[\boldsymbol{F}_{{\rm RF}}]_{:,k}^{(t-1)}\big\|_2^2\leq \epsilon$ or $t>T_{{\rm max}}$.

\section{Simulation Results}
\begin{figure}[t]
	\centering	
	\includegraphics[width=3.4 in]{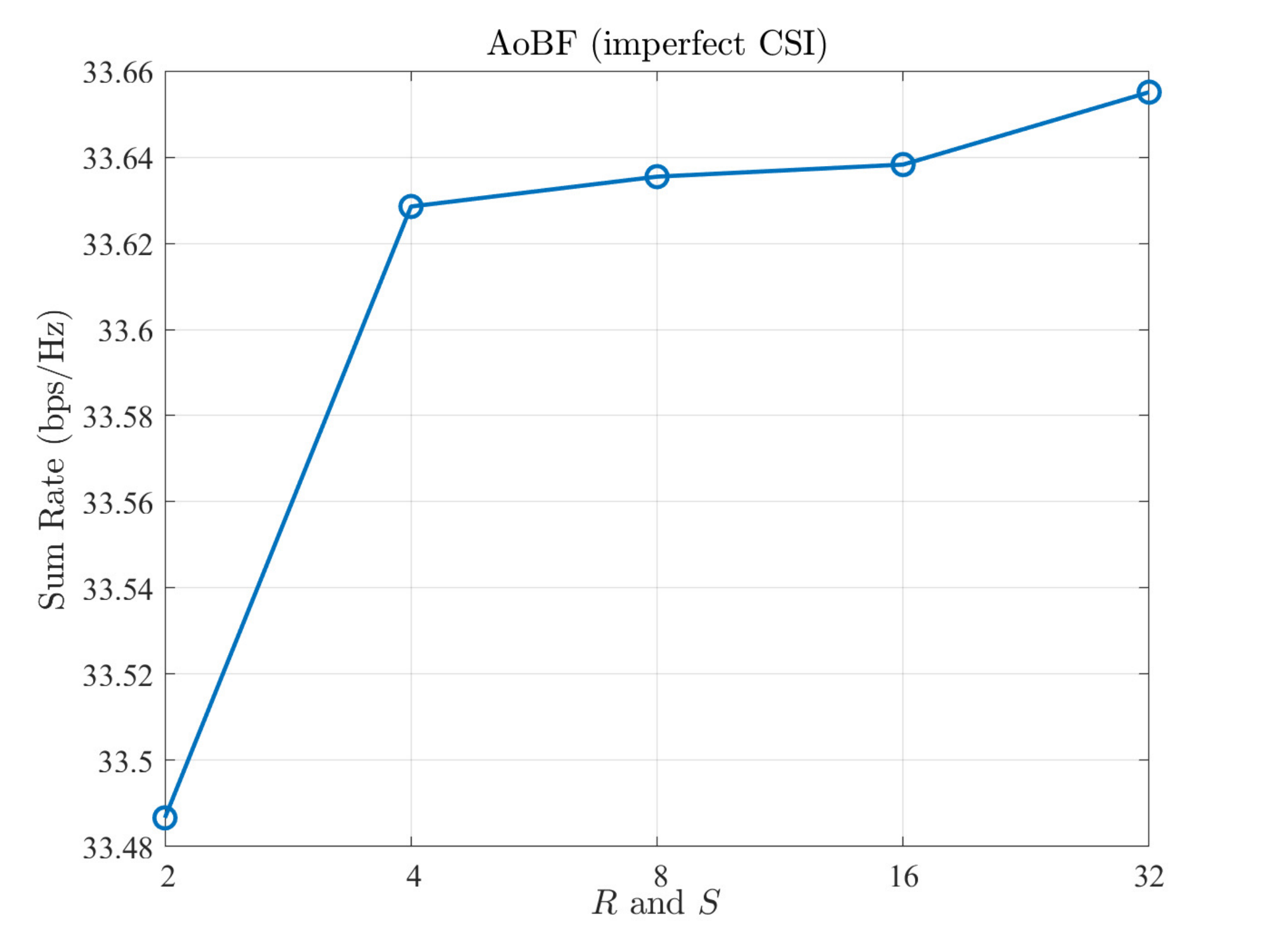}
	\caption{The performance of AoBF  based on imperfect CSI with different $R$ and $S$. }
	\label{fig9}
\end{figure}

In this section, the performance of the proposed AoBF schemes are evaluated. For the downlink transmission, a ULA with $N_{{\rm BS}}=64$ antennas is equipped at the BS.  The number of channel paths is set to be $L=3$ for each UE. The complex-valued channel gains follow $\alpha_{k,1} \sim \mathcal{C N}\left({0}, 1\right)$ and $\alpha_{k,l} \sim \mathcal{C N}\left({0}, 0.01\right)$ for $l=2,3$. There are $K=4$ UEs served by the BS, and the locations of UEs and scatterers are randomly generated within the near field. More specifically, the angles follow a uniform distribution in $[-\pi/2,\pi/2)$, and the distances to the central point of the ULA are randomly generated within the Rayleigh distance $D_{{\rm R}}$. The weight parameter of the MM method is set to be $\omega=1000$.  The parameters of the \textit{stop condition} in {\bf Algorithms~\ref{alg1} }and {\bf \ref{alg2}} are set to be $\epsilon =10^{-9}$ and $T_{{\rm max}}=1000$.  For the beam sweeping in the near field, the number of the code words in the angle dimension is set to be $N_{{\rm BS}}=64$, and the number of the code words in the distance dimension is set to be $N_{{\rm DIS}}=320$. The parameter of the polar-domain near-field codebook $\boldsymbol{\mathcal{V}}$ is set to be $\beta_{\Delta}=1.6$~\cite{ununi_codeword}. According to Fig.~\ref{fig9}, the changing point of the auxiliary point number is $4$. Thus, we set $R=4$ and  $S=4$. Monte Carlo simulations are performed based on 2000 random channel implementations.

\subsection{Beam Pattern Comparison}
\begin{table*}[h]\footnotesize
	\centering
	\caption{Normalized Beam Gain of Each UE} \label{Table:1}
	\addtolength{\tabcolsep}{4.8pt}
	\begin{tabular*}{5.5  in}{ccccc}
		\toprule[0.75pt]
		Schemes  &CSI & UE 1 &  UE 2  &  UE 3                  \\
		
		\midrule[0.5pt]
		Analog-only beam steering  &Perfect  &    $-7.5991e^{-7}$~dB   & -15.3376~dB &                     -14.6842~dB                      \\
		Analog-only beam steering  &Imperfect   &    -2.1414~dB    & -15.9440~dB &                     -15.2837~dB                    \\
		AoBF  & Perfect   &    -0.0105~dB    & -40.3274~dB &    -32.4819~dB                     \\
		AoBF  & Imperfect  &    -2.1255~dB    & -31.3642~dB &    -21.0754~dB                    \\
		HBF-WMMSE  & Perfect     &    -0.0130~dB    & -41.3534~dB &                     -37.2225~dB                     \\
		HBF-WMMSE & Imperfect    &    -2.4056~dB    & -36.2031~dB &                    -24.1222~dB                   \\
		\bottomrule[0.75pt]		
	\end{tabular*}
\end{table*}

\begin{figure*}[!t]
	\centering	
	\includegraphics[width=6.7 in]{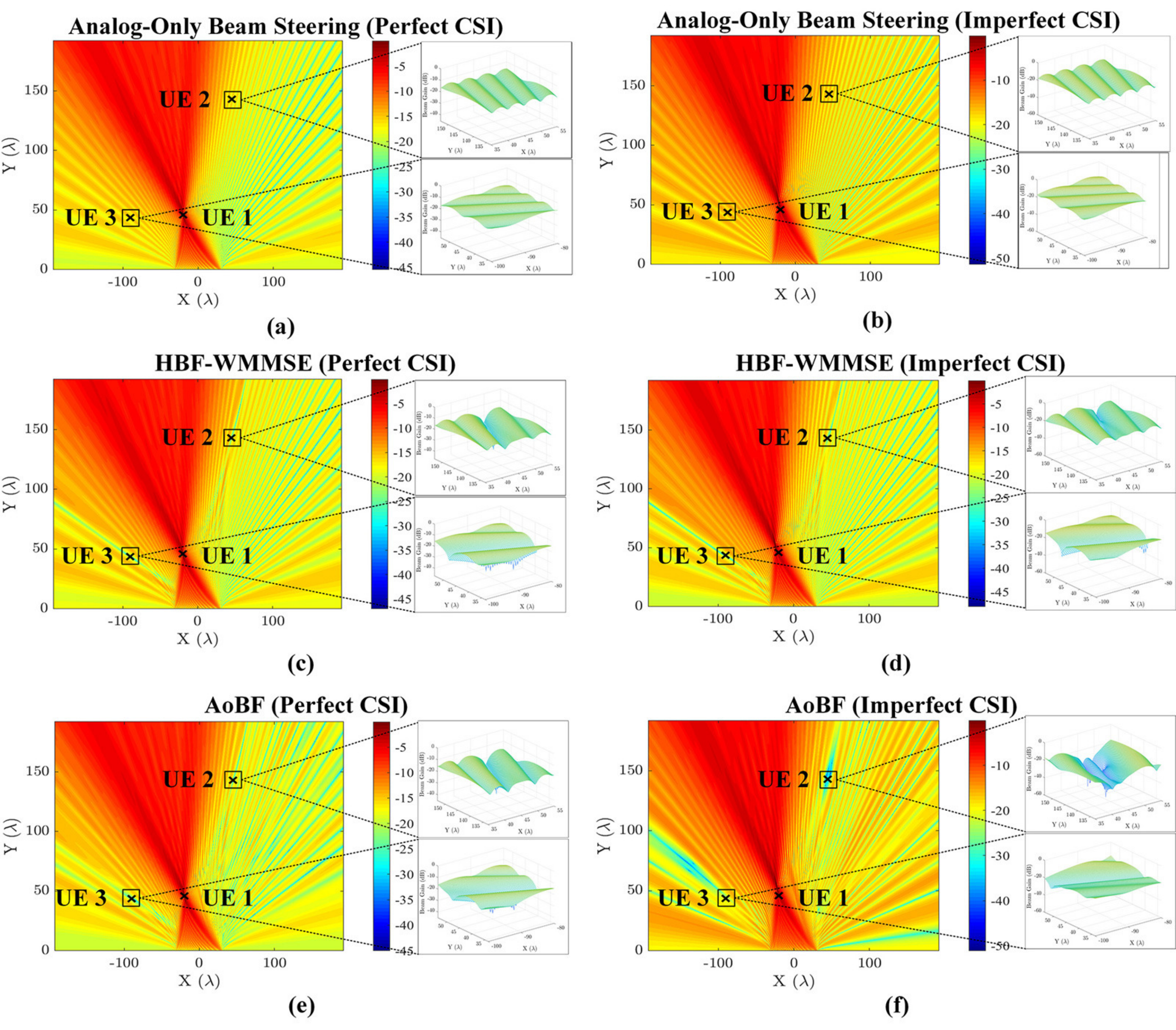}
	\caption{Beam pattern comparison: (a) the beam pattern generated by analog-only beam steering based on perfect CSI; (b) the beam pattern generated by analog-only beam steering based on imperfect CSI; (c) the beam pattern generated by HBF with WMMSE based on perfect CSI; (d) the beam pattern generated by HBF with WMMSE based on imperfect CSI; (e) the beam pattern generated by AoBF based on perfect CSI; (f) the beam pattern generated by AoBF based on imperfect CSI.}
	\label{fig5}
\end{figure*}
The beam patterns that are generated by analog-only beam steering, HBF with WMMSE-based digital beamforming, and our AoBF based on perfect CSI and imperfect CSI schemes are exhibited in Fig.~\ref{fig5}. There are $K=3$ UEs served by the BS. The locations of these UEs in polar coordinates are set to be $(-23.57^{\circ}, 50\lambda)$, $(17.46^{\circ}, 150\lambda)$, and $(-64.16^{\circ}, 100 \lambda)$,  respectively. We set the first UE located at $(-23.57^{\circ}, 50\lambda)$ as the target UE. As shown in Fig.~\ref{fig5}, the AoBF and HBF can mitigate the multiuser interference by maximizing the power transmitted to the target UE via beam focusing and minimizing the power leaked to the other UEs via beam nulling in the near field. The normalized beam gains  focused on the UEs are listed in Table~\ref{Table:1}. The beam gains from the BS to the target UE in the perfect-CSI-based schemes are close to $0~{\rm dB}$ since the beam can be accurately focused on the target UE with the help of the perfect CSI. On the contrary, the beam gains from the BS to the target UE in the imperfect-CSI-based schemes are lower to about $-2~{\rm dB}$ since the UE locations obtained by beam sweeping are inaccurate and the inaccurate beam focusing leads to gain loss. Based on perfect CSI, AoBF can achieve similar beam patterns as HBF. Based on imperfect CSI, owing to the generation of auxiliary points, the nearby power around the unimportant UEs is also attenuated in AoBF. Thus, it appears as a null-beam region. These results intuitively indicate the effectiveness of our AoBF schemes for multiuser interference suppression via beam focusing and beam nulling.

\subsection{Sum Rate Evaluation}

\begin{figure}[t]
	\centering	
	\includegraphics[width=3.4 in]{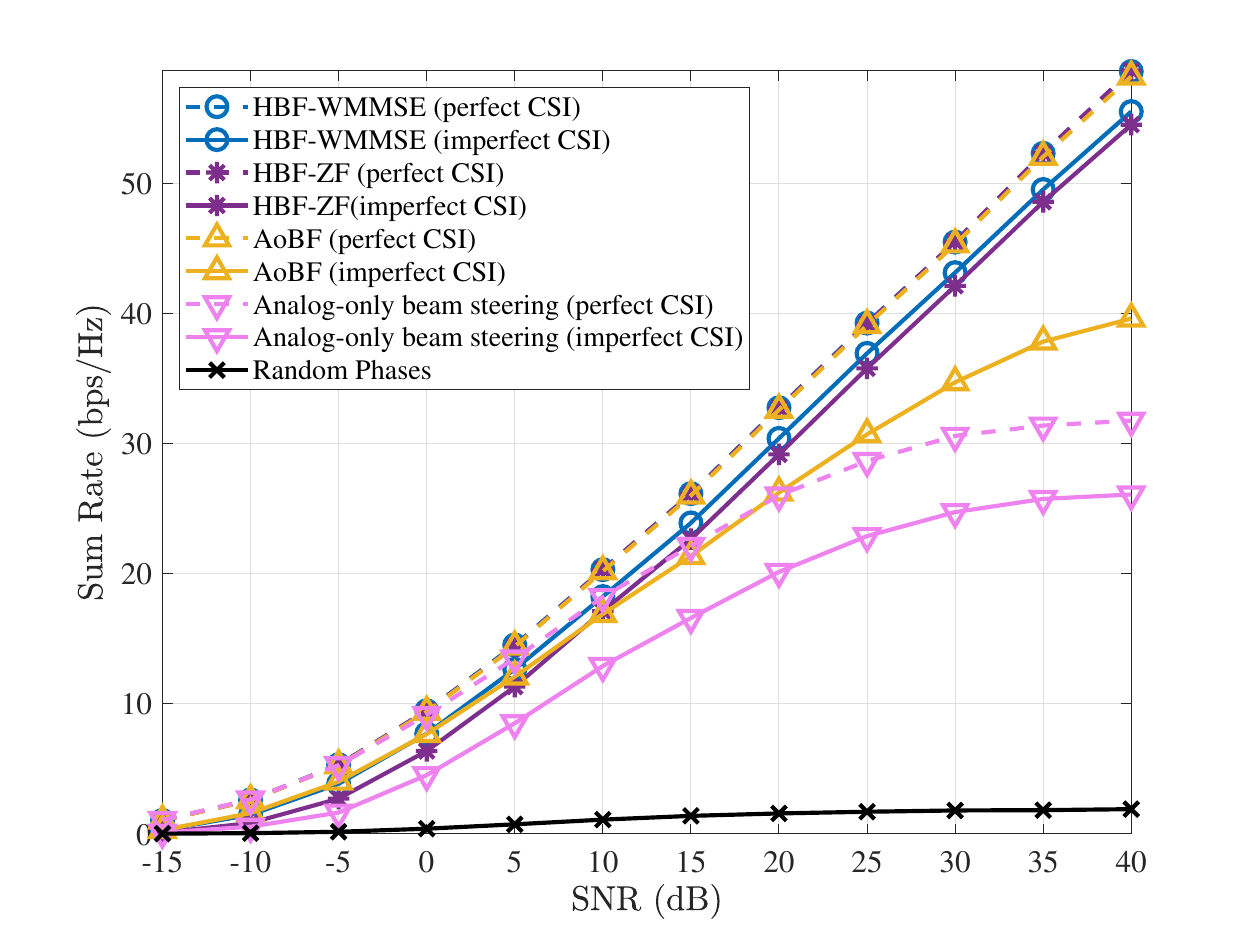}
	\caption{Comparisons of the proposed schemes with the existing schemes in
		terms of the sum rate for different SNRs.  }
	\label{fig6}
\end{figure}

In Fig.~\ref{fig6}, the performance of AoBF is compared with analog-only beam steering~\cite{Limited}, HBF with ZF-based digital beamforming, and HBF with WMMSE-based digital beamforming~\cite{Dai} in the near field, respectively.
{  The performance is evaluated in terms of the sum rate with different signal-to-noise ratios (SNRs). The SNR is defined as ${{P}{{\left| \boldsymbol{h}_k^{\rm H} \left[\boldsymbol{F}\right]_{:, k} \right|}^{2}}}/{(K{\sigma }^{2})}$. } The number of UEs is set to be $K=4$. Particularly, the perfect-CSI-based schemes can be considered to show the performance of upper bounds. Based on the perfect CSI, the performance of our AoBF schemes can approach that of HBF-ZF and HBF-WMMSE. This implies the possibility of replacing HBF with AoBF. Based on imperfect CSI, HBF-WMMSE and HBF-ZF outperform AoBF since digital beamforming in HBF provides more degree of freedom for beamforming compared with AoBF. Owing to the multiuser interference suppression ability of our AoBF scheme as shown in Fig.~\ref{fig5}b and f, AoBF outperforms analog beam steering. For instance, based on imperfect CSI, AoBF can achieve $29.18\%$ performance improvement over the analog-only beam steering at SNR $=20$~dB.

\begin{figure}[t]
	\centering	
	\includegraphics[width=3.4 in]{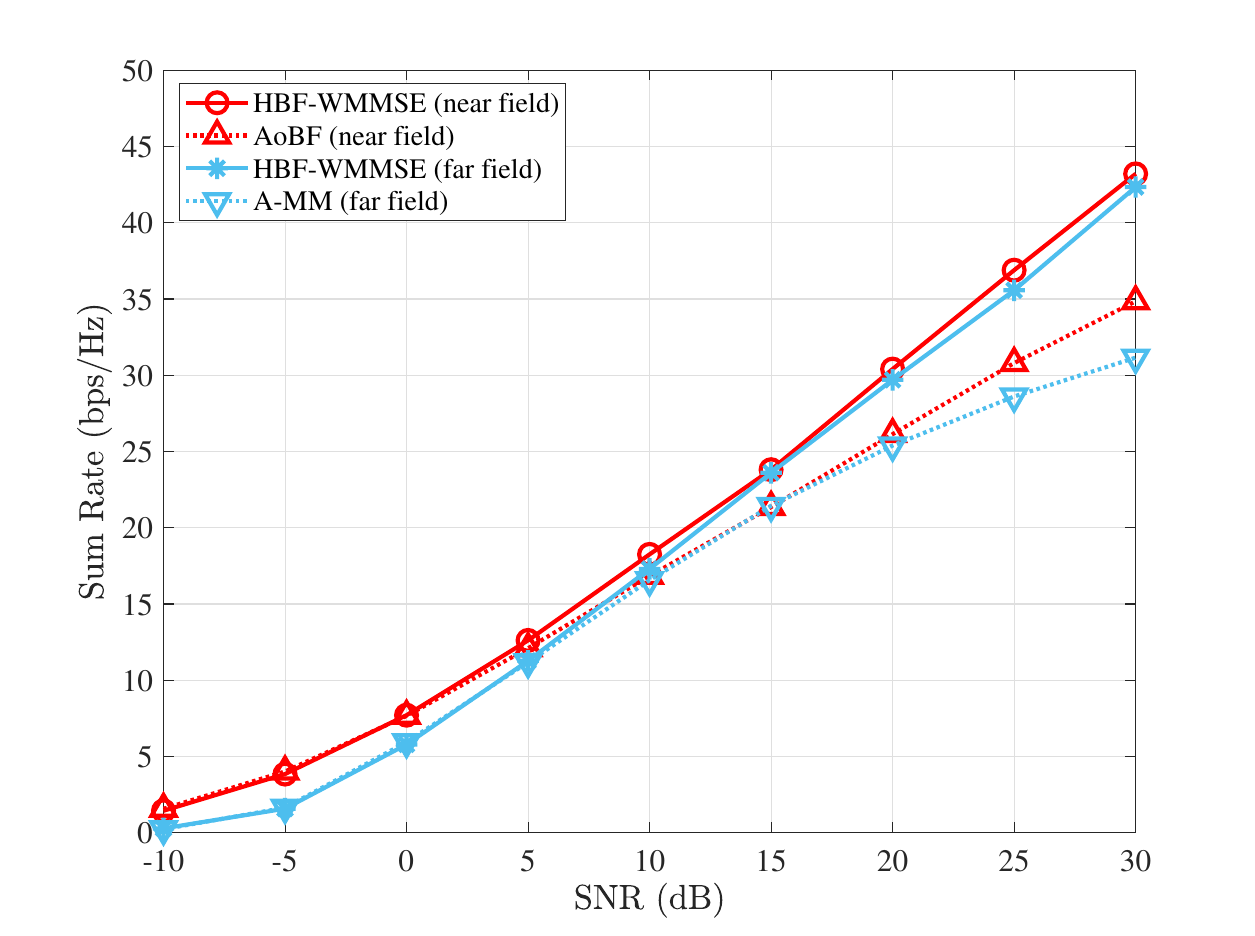}
	\caption{Comparisons of the near-field schemes and the far-field schemes.}
	\label{fig10}		
\end{figure}
\begin{figure}[t]
	\centering	
	\includegraphics[width=3.4 in]{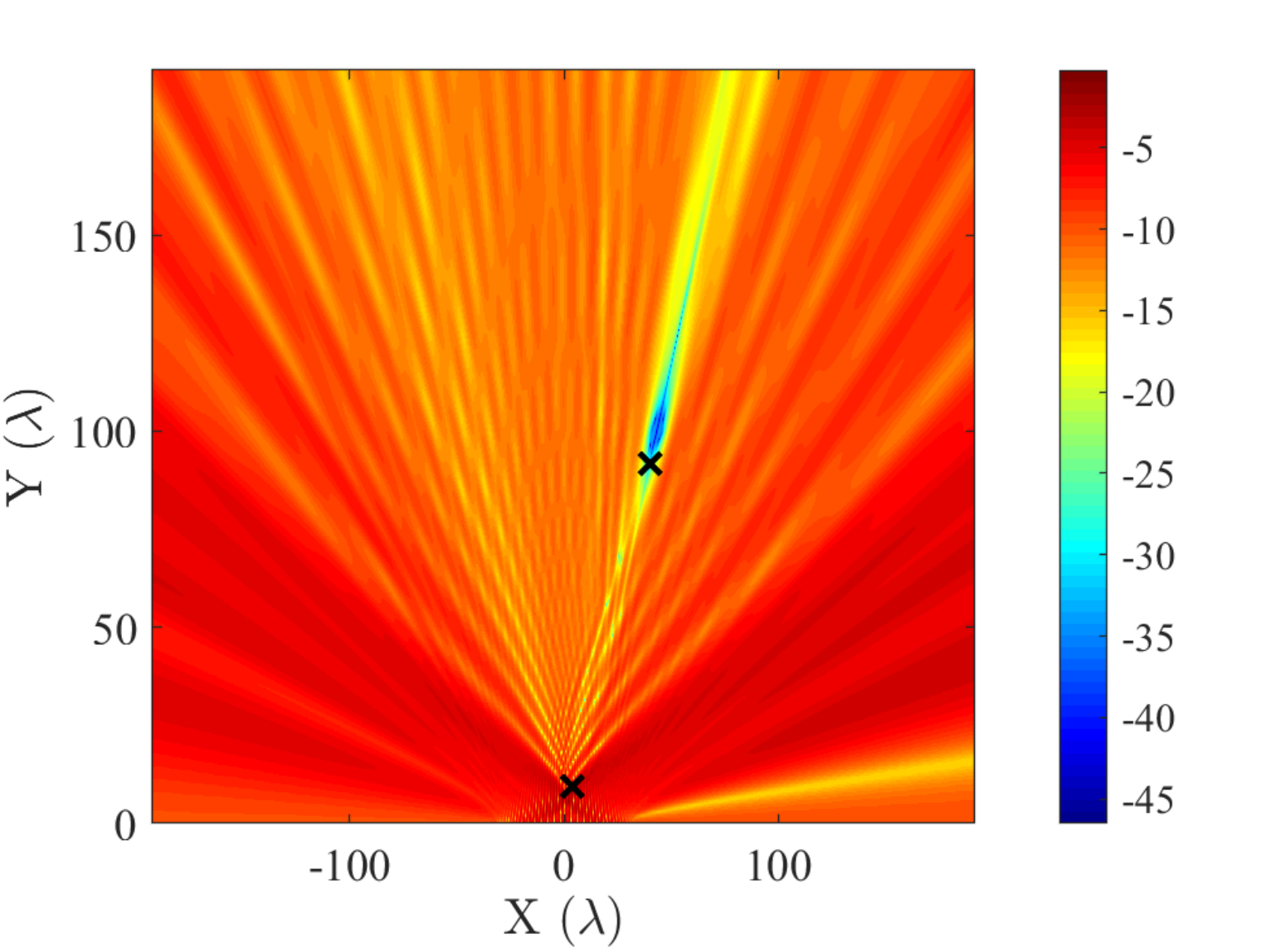}
	\caption{Illustration of the beam pattern generated by imperfect-CSI-based AoBF if the UEs are located at the same angle in the near field.}
	\label{fig11}
	
\end{figure}
In Fig.~\ref{fig10}, the performance of AoBF in the near field is compared with the existing far-field schemes, including the far-field HBF-WMMSE scheme~\cite{Limited} and the far-field A-MM scheme~\cite{HJL}. { For a fair comparison, we extend the partially connected A-MM scheme~\cite{HJL} to the fully connected structure. The A-MM scheme is designed based on far-field modeling and a far-field codebook. Thus, it cannot distinguish UEs located at different distances in the near field. Our AoBF scheme is designed based on the near-field system model and codebook. The beamforming method is also adjusted accordingly. The ability to distinguish UEs in AoBF is improved, thanks to the introduction of the distance dimension.} As shown in Fig.~\ref{fig11}, AoBF can distinguish UEs located at the same angle and different distances. Thus, it can be seen in Fig.~\ref{fig10} that the near-field AoBF schemes perform slightly better than the far-field A-MM. The gap between AoBF and HBF-WMMSE in the near field is narrower than that between A-MM and HBF-WMMSE in the far field. At SNR $=20$~dB, AoBF achieves $85.90\%$ of the sum rate of HBF-WMMSE in the near field, while A-MM can only achieve $85.50\%$ of the sum rate of HBF-WMMSE in the far field. Compared with the far-field A-MM, the better performance of near-field AoBF is achieved due to the extra degree of freedom in the distance dimension.

\begin{figure}[!t]
	\centering	
	\includegraphics[width=3.4 in]{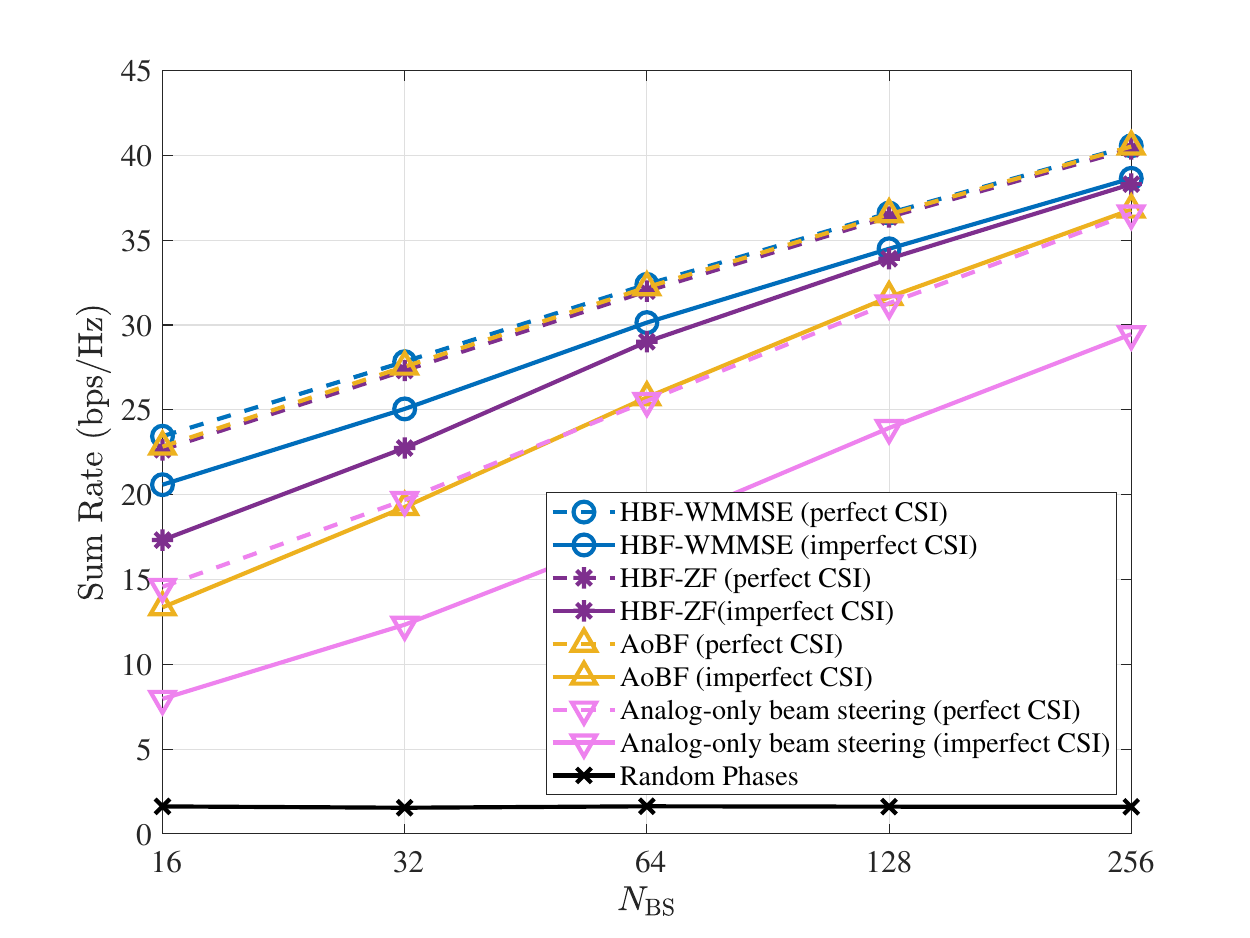}
	\caption{Comparisons of the proposed schemes with the existing schemes in terms of the sum rate for different $N_{{\rm BS}}$.}
	\label{fig7}
	
\end{figure}

In Fig.~\ref{fig7}, we compare AoBF with the existing schemes in terms of the sum rate with different $N_{{\rm BS}}$. In this simulation, the UE number is set to be $K=4$, and the SNR is set to be $20$~dB. It can be seen that the performance of all schemes improves with the increase of $N_{{\rm BS}}$. This is caused by two reasons. First, the ULA with more antennas generates more focused beams and, therefore, achieves larger beam gains. Second, the resolution of beam sweeping with a larger array is higher. The accuracy of the UE location from beam sweeping is improved, and the gap between the imperfect CSI and the perfect CSI is reduced with the increase of $N_{{\rm BS}}$. Based on perfect CSI, the performance of AoBF and HBF is close. Thus, the gap between the AoBF and the HBF schemes based on imperfect CSI becomes narrower with the increase of $N_{{\rm BS}}$. Therefore, the increase of the antenna number is an effective way to improve the performance of AoBF if the cost is permitted.

\begin{figure}[t]
	\centering	
	\begin{tabular}{c}
		\includegraphics[width=3.4 in]{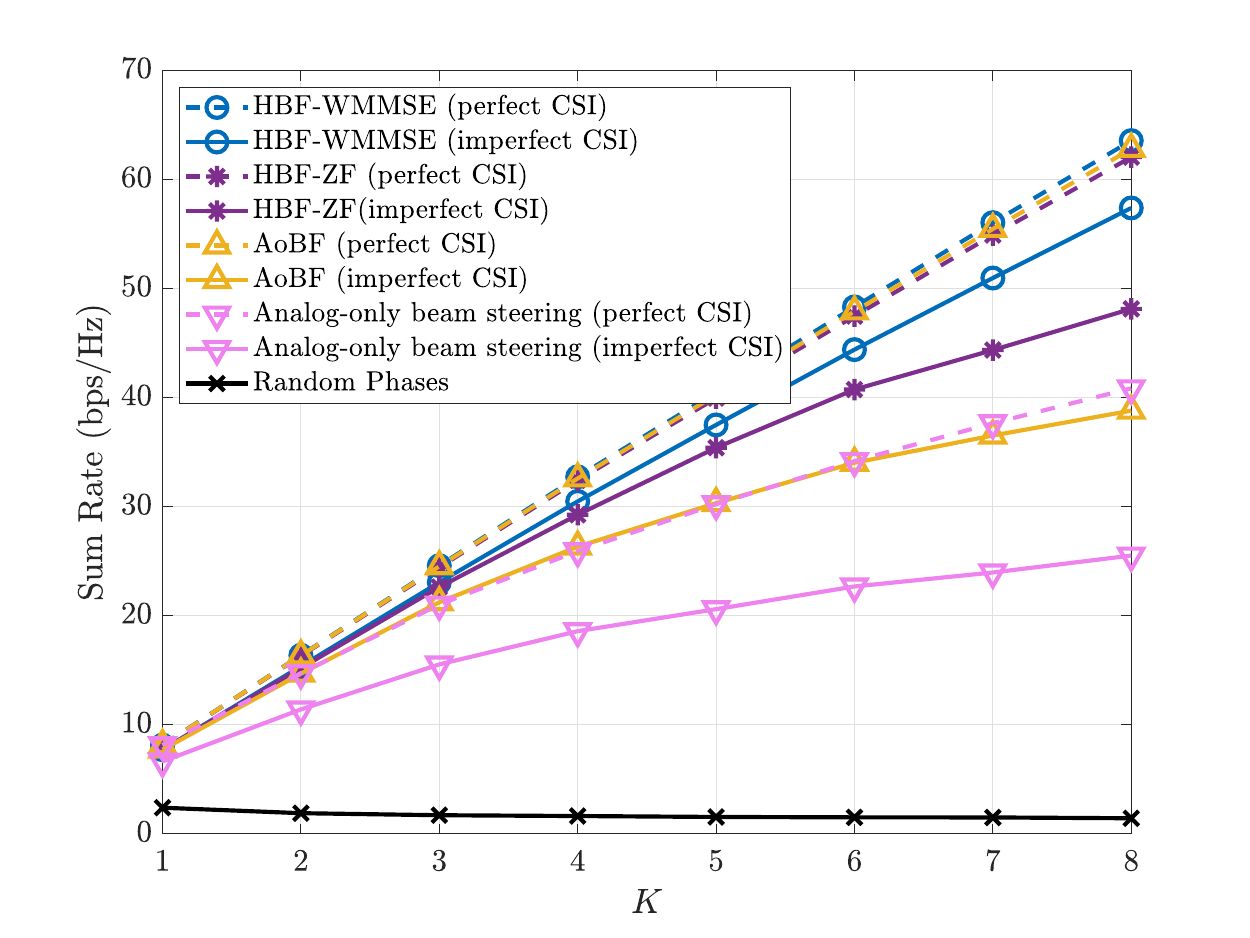}\\
		{\footnotesize\sf (a)} \\[3mm]
		\includegraphics[width=3.4 in]{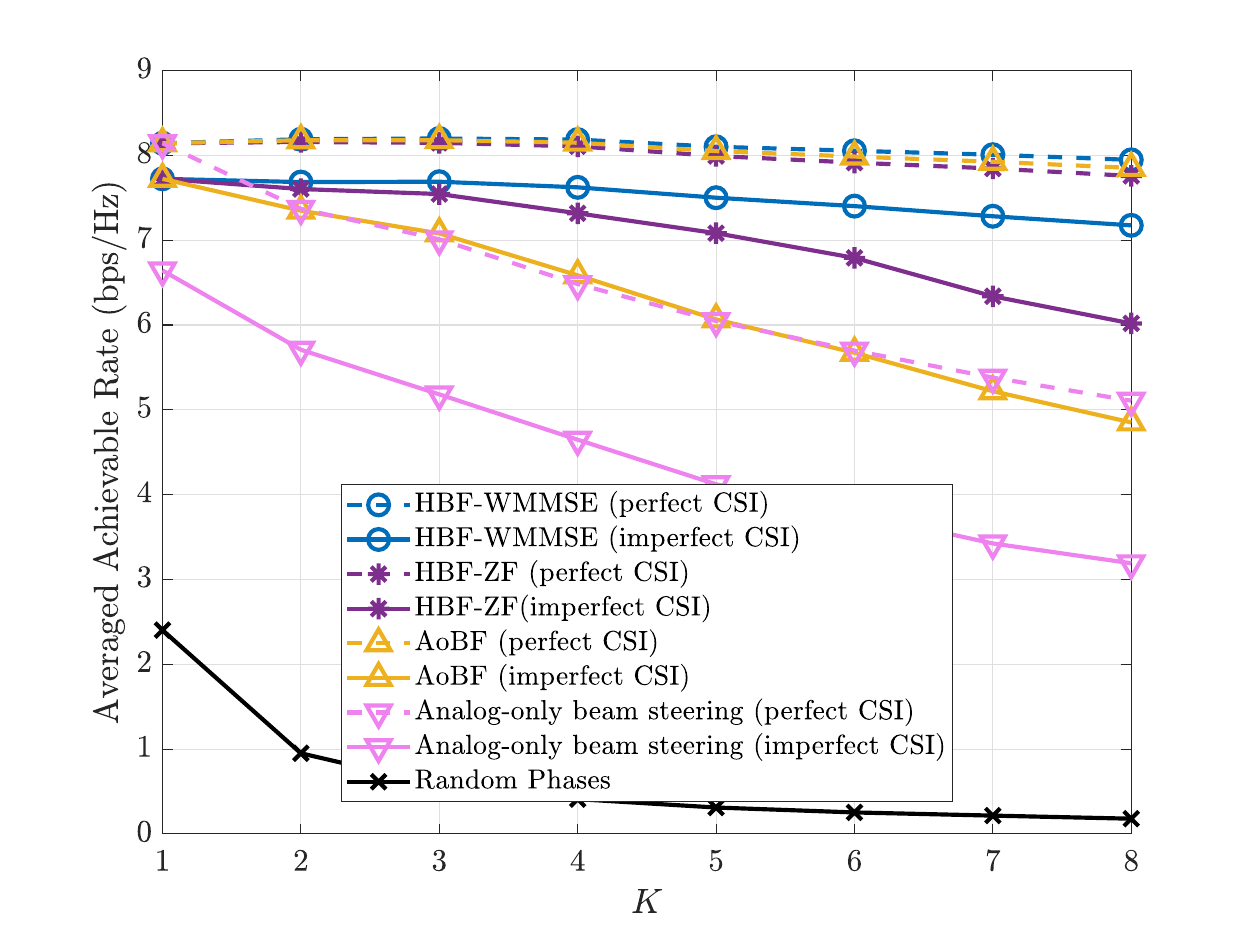}\\
		{\footnotesize\sf (b)} \\
	\end{tabular}
	\caption{Comparisons of the proposed schemes with the existing schemes for different $K$: (a) in terms of the sum rate; (b) in terms of the averaged achievable rate.}
	\label{fig8}
	
\end{figure}

In Fig.~\ref{fig8}, we compare AoBF with the existing schemes in terms of the sum rate and the averaged achievable rate $(\sum_{k=1}^K R_k)/K$ with different $K$. In this simulation, the antenna number is set to be $N_{\rm BS}=64$, and the SNR is set to be $20$~dB. It can be seen that the sum rate of all schemes increase with the increase of $K$. The averaged achievable rates of these schemes decrease with the increase of $K$. The reason is that the interference among UEs becomes more severe with larger $K$.

\subsection{EE Evaluation}
Compared with HBF, AoBF removes the digital beamforming module and, therefore, can achieve system reconfiguration simplification and EE improvement. The EE is defined as follows~\cite{EE1, BB2}: 
\begin{equation}
	\label{eq23}
	{\rm EE}\triangleq\frac{\sum\limits_{k=1}^K {R_k}}{P_{\rm total}},
\end{equation}
where
\begin{equation}
	\label{eq24}
	\begin{aligned}
		P_{\rm total}\triangleq&P+N_{\rm RF}P_{\rm RF}+N_{\rm BS}N_{\rm RF}P_{\rm PS}+P_{\rm BB}.
	\end{aligned}
\end{equation}
The variables $P$, $P_{\rm RF}$, $P_{\rm PS}$, and $P_{\rm BB}$ represent the power of the BS, each RF chain, each phase shifter, and digital beamforming in the baseband, respectively.  We set $P_{\rm RF}=26~{\rm mW}$, $P_{\rm PS}=10~{\rm mW}$, and $P_{\rm BB}=200~{\rm mW}$~\cite{BB1,BB3}.
\begin{figure}[t]
	\centering	
	\includegraphics[width=3.4 in]{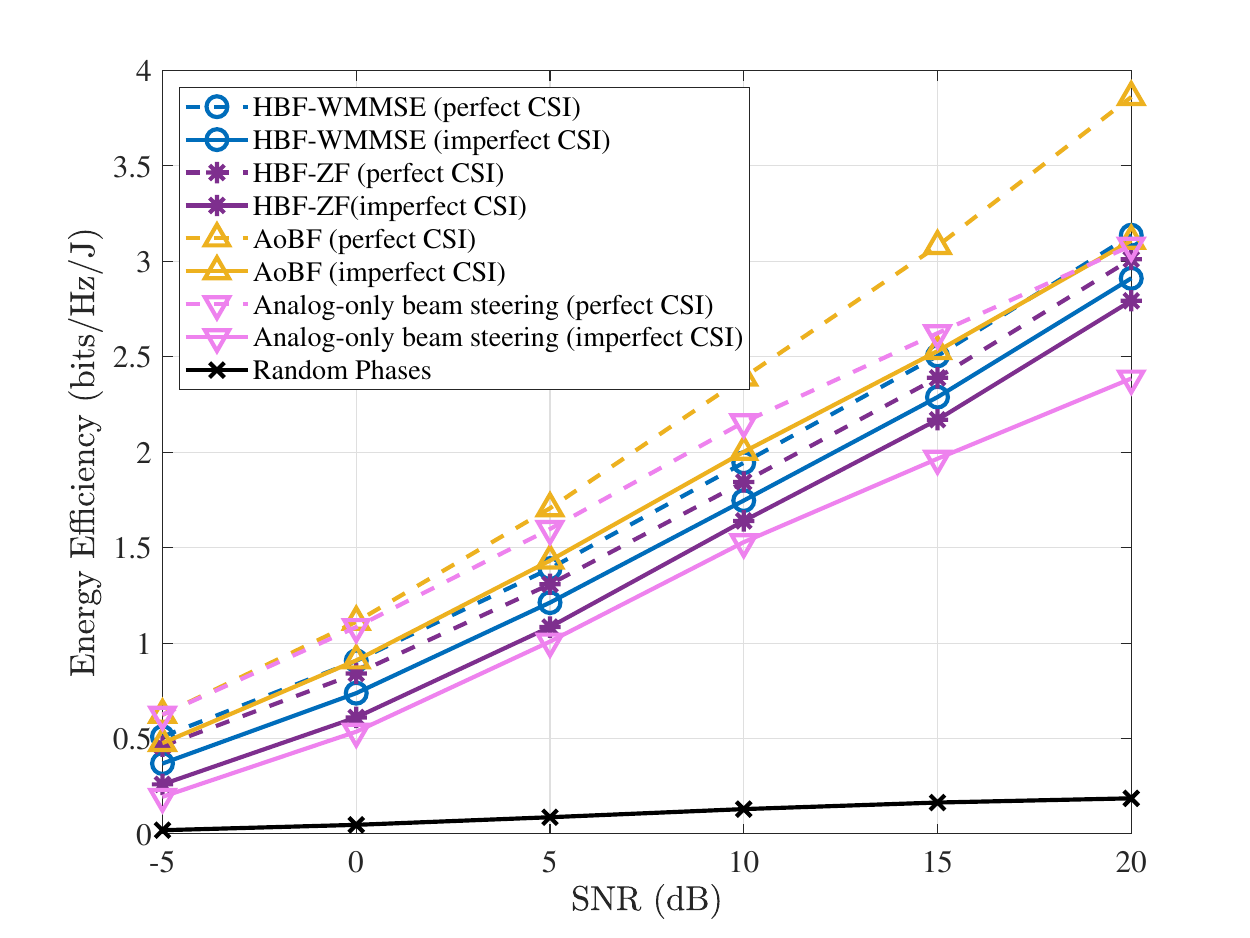}
	\caption{Comparisons of the proposed schemes with the existing schemes in
		terms of EE for different SNRs.}
	\label{fig4}
\end{figure}

As shown in Fig.~\ref{fig4}, we compare AoBF with the existing schemes in terms of EE with different SNRs. Our AoBF schemes achieve better EE than the HBF schemes~\cite{Dai}, owing to the removal of the digital beamforming module. In addition, AoBF outperforms analog-only beam steering due to the higher sum rate achieved by multiuser interference suppression in AoBF. Based on imperfect CSI, AoBF achieves $14.60\%$ and $30.95\%$ EE improvement over the HBF-WMMSE and the analog-only beam steering at SNR $=10$~dB, respectively.

\section{Conclusion}
In this study, we have proposed two AoBF schemes for multiuser near-field communications based on perfect CSI and imperfect CSI, respectively. The proposed AoBF schemes can approach the performance of HBF with simplified system configuration and higher EE. The future study will focus on the AoBF design based on PCM.
	
\bibliographystyle{IEEEtran}
\bibliography{bibsample}
	
	\vfill
	
\end{document}